\documentclass[twocolumn,prd,superscriptaddress,showpacs,floatfix,%  
preprintnumbers, nofootinbib,hyperref]{revtex4}  
\usepackage{epsfig}   
\usepackage{bm}  
\usepackage{color}

\newcommand{\ben}{\begin{equation}}
\newcommand{\een}{\end{equation}}

\newcommand{\gtrsim}{\,\rlap{\lower3.7pt\hbox{$\mathchar\sim$}}
\raise1pt\hbox{$>$}\,}
\newcommand{\lesssim}{\,\rlap{\lower3.7pt\hbox{$\mathchar\sim$}}
\raise1pt\hbox{$<$}\,}

\newcommand{\be}{\begin{equation}}
\newcommand{\ee}{\end{equation}}  
\newcommand{\bea}{\begin{eqnarray}}
\newcommand{\eea}{\end{eqnarray}}

\begin{document}

\title{No collective neutrino flavor conversions during the supernova accretion phase}

\author{Sovan Chakraborty} 
\affiliation{II Institut f\"ur Theoretische Physik, Universit\"at Hamburg,
Luruper Chaussee 149, 22761 Hamburg, Germany}
 
\author{Tobias Fischer} 
\affiliation{GSI, Helmholtzzentrum f\"ur Schwerionenforschung GmbH,
Planckstra{\ss}e 1, 64291 Darmstadt, Germany}

\affiliation{Technische Universit\"at Darmstadt, Schlossgartenstra{\ss}e 9,
64289 Darmstadt, Germany}

\author{Alessandro Mirizzi} 
\affiliation{II Institut f\"ur Theoretische Physik, Universit\"at Hamburg,
Luruper Chaussee 149, 22761 Hamburg, Germany} 

\author{Ninetta Saviano} 
\affiliation{II Institut f\"ur Theoretische Physik, Universit\"at Hamburg,
Luruper Chaussee 149, 22761 Hamburg, Germany}  
 
\author{Ricard Tom{\`a}s}   
\affiliation{II Institut f\"ur Theoretische Physik, Universit\"at Hamburg,
Luruper Chaussee 149, 22761 Hamburg, Germany}

\begin{abstract}
We perform a dedicated study of the SN neutrino flavor evolution during the
accretion phase,  using results from  recent neutrino
radiation hydrodynamics simulations.
In contrast to what expected in the presence of only neutrino-neutrino interactions, we find that the multi-angle effects associated
with the dense ordinary matter suppress collective oscillations.
The matter suppression implies that neutrino oscillations will start outside the
 neutrino decoupling region and therefore will have a
negligible impact on the neutrino heating and the explosion dynamics.
Furthermore, the  possible detection of the next
galactic SN neutrino signal from the accretion phase, based on the
usual Mikheyev-Smirnov-Wolfenstein effect in the SN mantle
and  Earth matter  effects,
can reveal the neutrino mass hierarchy in the  case that the
mixing angle $\theta_{13}$ is not very small.
\end{abstract}


\maketitle

\emph{Introduction.---}
Neutrinos emitted  from  core-collapse supernovae (SNe)
represent a crucial tool to get valuable information about the  mixing parameters
and an insight into the dynamics of the exploding stellar core~\cite{Raffelt:2010zza}.
SN  neutrinos not only  interact with the stellar medium
via the Mikheyev-Smirnov-Wolfenstein (MSW) effect~\cite{Matt},  but also with other
 neutrinos  ($\nu$) and antineutrinos 
(${\overline\nu}$).
It was pointed out that large $\nu$ densities in the deepest stellar regions can result
in significant coherent $\nu$--$\nu$ forward
scatterings~\cite{Pantaleone:1992eq,Qian:1994wh},
which give rise to collective $\nu$ flavor oscillations inside the
SN~\cite{Duan:2005cp,Duan:2006an,Hannestad:2006nj}
(see~\cite{Duan:2010bg} for a recent review).

The development of these self-induced $\nu$ transformations crucially depends
on the primary SN $\nu$ spectra
(see, e.g.,~\cite{Fogli:2007bk,Dasgupta:2009mg}). 
At this regard the post-bounce accretion phase of core-collapse SNe, lasting few
tens  of milliseconds (for low mass O-Ne-Mg-core progenitors)
up to several hundreds of milliseconds (for more massive iron-core progenitors),
might seem the best opportunity to detect signatures of collective
$\nu$ flavor oscillations.
Indeed, the absolute $\nu$ fluxes are  large during the accretion phase
with  significant spectral differences between the different $\nu$
species, and a flux order
$F_{\nu_e} > F_{{\overline\nu}_e} \gg F_{\nu_x}$. 
This scenario has been often  taken as a benchmark for the description
of the self-induced effects.
Notably, these latter would leave the $\nu$ spectra unaffected  in normal mass hierarchy
(NH: $\Delta m^2_{\rm atm} = m_3^2 - m_{1,2}^2>0$).
In the inverted $\nu$ mass hierarchy (IH: $\Delta m^2_{\rm atm}<0$),
they would produce a complete exchange of the $\bar{\nu}_e$ and
$\bar{\nu}_x$ spectra, and a spectral split in the energy distributions of the
 $\nu$'s~\cite{Fogli:2007bk}.
This seemingly robust and clear behavior has been proposed as an unique
way to determine the $\nu$ mass hierarchy even if the leptonic 1--3 
mixing angle $\theta_{13}$ is too small to be detected in terrestrial $\nu$
oscillation experiments~\cite{Duan:2007bt,Dasgupta:2008my}. 

The implicit assumption in this picture  is related to the flavor evolution
in the deepest SN regions being driven by only large neutrino densities
$n_{\nu}$.
However, during  the accretion phase also the  net 
electron density $n_e$ is expected to be large, as documented by many different SN
simulations~\cite{shock,Tomas:2004gr,Buras:2005rp, Liebendoerfer:2003es}. 
This is a generic feature that applies to  SNe of massive iron-core
progenitors.
As recently pointed out in~\cite{EstebanPretel:2008ni} and confirmed in~\cite{Duan:2008fd}, when $n_e$
is not negligible with respect to $n_\nu$, the large phase dispersion induced by the
matter for $\nu$'s traveling in different directions, will  partially
or totally suppress the collective oscillations through peculiar multi-angle effects.

Motivated by this  insight,  we have performed
a  detailed  study of the SN $\nu$ flavor evolution
during the accretion phase, characterizing the $\nu$ signal and the matter density
profiles  by means of recent neutrino radiation  hydrodynamics simulations.
Contrarily to what  shown in previous studies based on the only $\nu$-$\nu$ interaction effects,  
we find that the presence of a dominant matter term inhibits the development
of collective flavor conversions.
The matter suppression  ranges  from complete to partial,
producing intriguing time-dependent features.
In particular, when it is complete (for post-bounce times $t_{\rm pb} \lesssim 0.2$~s in
iron-core SNe) the $\nu$ signal will be processed only by the usual
MSW effect in the SN mantle and Earth matter  effects.
This was the usual description before the inclusion of collective 
phenomena. 
This \emph{d\'ej{\`a} vu} would reopen the possibility, prevented by self-induced effects,
to reveal the neutrino mass hierarchy through the Earth matter effect on the next galactic
SN neutrino burst, in the case  $\theta_{13}$ is not very
small~\cite{Dighe:2003jg}.

\begin{figure}[!t]
\begin{center}  
\includegraphics[width=.75\columnwidth]{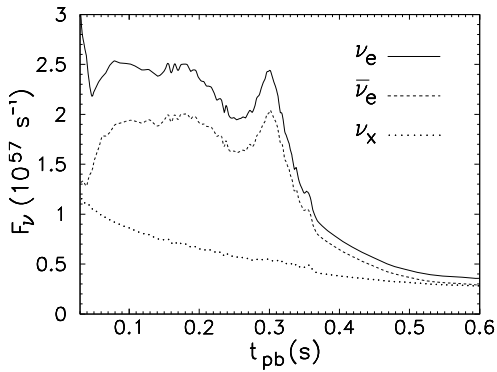}  
\end{center} 
\vspace{-0.5cm} 
\caption{$\nu$ fluxes during the accretion phase of the 10.8~M$_\odot$ SN
explosion model.
\label{fig:1}}  
\end{figure}

\emph{$\nu$ signal from the accretion phase.---}
We take as benchmark for our study the results of the recent long-term SN
simulations, described in~\cite{Fischer:2009af}.
These are based on radiation hydrodynamics that employs three flavor Boltzmann
neutrino transport in spherical symmetry.
Figure~\ref{fig:1} shows the evolution of the $\nu$ number fluxes $F_{\nu_\alpha}$
for the different neutrino flavors $\nu_\alpha$ up to 0.6~seconds after core bounce,
for the 10.8~M$_\odot$ iron-core progenitor model. Enhanced $\nu$ heating was applied, because
a neutrino-driven explosion cannot be obtained in spherical symmetry for such a progenitor
(for details, see~\cite{Fischer:2009af}). 
The first phase after core bounce lasts only $\sim 0.02$~s,
where large numbers of electron captures release a flare of $\nu_e$
with luminosities on the order of $10^{53}$~erg/s.
It is followed by the accretion phase that can last up to several
hundred milliseconds.
After the onset of the explosion, mass accretion vanishes at the neutrinospheres
(i.e., $\nu$ last-scattering surfaces) and the neutrino luminosities are determined
by diffusion.
It results in a sharp drop of the fluxes after the explosion shock crosses a distance
of  500~km, where the fluxes are measured in a co-moving reference frame.
The fluxes of all flavors decrease continuously on a longer timescale of
${\mathcal O}(10$~s), indicating the beginning of the cooling phase.

\emph{Setup of the flavor evolution.---}
Our description of the  $\nu$ flavor conversions is based on a
two-flavor scenario,
 driven by the atmospheric mass-square difference
$\Delta m^2_{\rm atm} \simeq 2.6 \times 10^{-3}$~eV$^{2}$
and by a small (matter suppressed) in-medium mixing
$\theta_{\rm eff} = 10^{-3}$~\cite{Schwetz:2011qt}.
Three-flavor effects, associated with the solar sector, are small for the $\nu$
flux ordering expected during the accretion phase~\cite{Mirizzi:2010uz}.
We will always refer to the inverted mass hierarchy where for the assumed
spectral ordering collective oscillations  are possible~\cite{Fogli:2007bk,Mirizzi:2010uz}.
The impact of the non-isotropic nature of the $\nu$ emission on the flavor conversions
is taken into account by ``multi-angle''  simulations~\cite{Duan:2006an}, where one
follows a large number $[{\mathcal O}(10^3)]$ of interacting $\nu$ modes.
The $\nu$'s emitted from a SN core naturally have
a broad energy distribution. However, this is largely irrelevant for our purposes, since large matter effects would lock togheter the 
different neutrino energy modes, both in case of  supression~\cite{EstebanPretel:2008ni}  and
of   decoherence~\cite{EstebanPretel:2007ec} of collective oscillations. 
 Therefore, to simplify the complexity of the numerical simulations, we assume all $\nu$'s to
be represented by a single energy, that we take $E=15$~MeV.

%%%%%%%%%%%%%%%%%%%%%%%%%%%%%%%%%%%%%%%%%%  
%%%%%%%%%%%%%%%%%%%%%%%%%%%%%%%%%%%%%%%%%%

The strength of the $\nu$--$\nu$ interaction is given
by
 %.................................................................
$
 \mu_r = \sqrt{2}G_F \left[n_{{\overline\nu}_e} (r) - n_{{\overline\nu}_x} (r) \right]
%\label{eq:mur}
 $~\cite{EstebanPretel:2007ec},
 %..................................................................
where $n_{\nu_\alpha}(r) = {F_{\nu_\alpha}}/{4\pi r^2} $
 %.......................................................
is the number density  of the $\nu$ species $\nu_\alpha$. 
The $\nu$--$\nu$ potential is normalized at the neutrinosphere, where $\nu$'s are
assumed to be half-isotropically emitted~\cite{EstebanPretel:2007ec}. 
We determine the neutrinospheres  from the core-collapse
SN simulations, where during the accretion phase the neutrinosphere radius is at
$r_{\nu} \sim {\mathcal O}(10^2)$~km.  
The matter potential is represented by
%...........................................................
$
\lambda_r = \sqrt{2}G_F  n_{e}(r)
$~\cite{Dighe:1999bi}, 
%.............................................................
encoding the net electron density, $n_e\equiv n_{e^-}- n_{e^+}$ .
 
%%%%%%%%%%%%%%%%%%%%%%%%%%%%%%%%%%%%%%%%%%
\begin{figure*}  
\includegraphics[angle=0,width=0.6\textwidth]{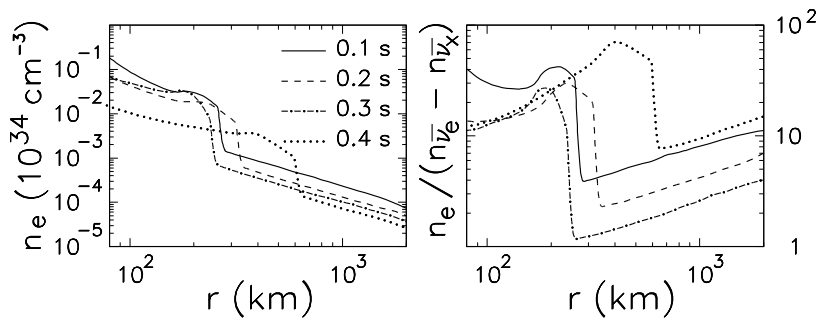}
\vspace{-0.4cm}   
\caption{
Radial profiles of the net electron density $n_e$ (left panel)
and of the ratio $R=n_e/(n_{\overline\nu_e}-n_{\overline\nu_x})$
(right panel), at selected post-bounce times for the
10.8~M$_\odot$ SN model.
\label{fig2}} 
\end{figure*}  
%%%%%%%%%%%%%%%%%%%%%%%%%%%%%%%%%%%%%%%%%%

\emph{Matter and neutrino densities.---}
In Fig.~\ref{fig2} we  show the net electron density $n_e$ (left panel)
and the ratio $R=n_e/(n_{\overline\nu_e}-n_{\overline\nu_x})$ between electron
and neutrino densities entering the potentials 
(right panel), at selected post-bounce times. 
One can recognize the abrupt discontinuity in $n_e$ associated with 
the SN shock-front that propagates in time.
From the ratio $R$ at the different post-bounce times, we realize that $n_e$ is
always larger than or comparable to $n_{\overline\nu_e}-n_{\overline\nu_x}$,
suggesting that matter effects cannot be ignored during the accretion phase.
Depending on the strength of the  matter density,  the
matter suppression  can  be total, when
$n_e\gg n_\nu$, or partial when the matter dominance is less pronounced.
Finally, when $n_e \gtrsim n_\nu$ the interference of the two comparable potentials
leads to a flavor equilibrium  with a complete
mixture of electron and non-electron species~\cite{EstebanPretel:2008ni}.
 
\begin{figure}[!t]
\begin{center}  
\includegraphics[width=.8\columnwidth]{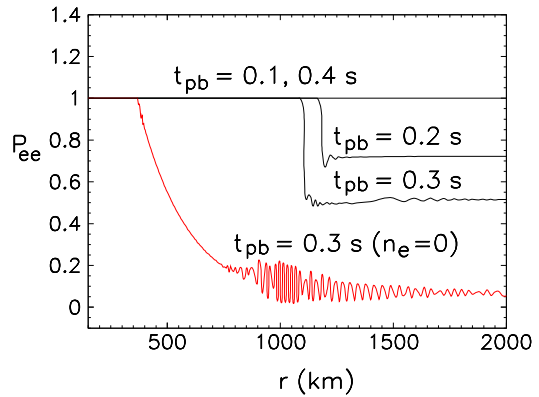}  
\end{center}  
\vspace{-0.5cm} 
\caption{Radial  profiles  of the
${\overline\nu}_e$ survival probability $P_{ee}$
at  selected  post-bounce times from
multi-angle simulations in matter (black continuous curves)
and for $n_e=0$ (light  curve). 
\label{fig3}} 
\end{figure}

\emph{Neutrino flavor conversions.---}
In order to have a quantitative description of these matter effects,
we performed a multi-angle numerical study of the $\nu$ flavor evolution
in the schematic model described above. 
In Fig.~\ref{fig3} we show the radial evolution of the ${\bar\nu}_e$ survival
probability $P_{ee}$ for the  same post-bounce times  as in 
Fig.~\ref{fig2}.
For comparison, we also show the example of what is expected in the case
of $n_e=0$ (light curve for $t_{\rm pb}=0.3$~s).
As predicted, we find that matter strongly suppresses
the development of the self-induced flavor transformations.  
In particular, at $t_{\rm pb}=0.1, 0.4$~s, when $n_e \gg n_{\nu}$
the flavor conversions are completely blocked ($P_{ee}=1$). 
Conversely, at $t_{\rm pb}=0.2$~s when $n_e \simeq 2 n_\nu$ in the conversions
region, the matter suppression  is only partial giving  a
final $P_{ee} \simeq 0.75$.  
Finally, at $t_{\rm pb}=0.3$~s when $n_e \gtrsim n_{\nu}$, matter effects
produce a complete flavor mixture ($P_{ee} = 1/2$).
From a systematic study of the flavor evolution at different time snapshots during
the accretion phase, we find
 {\em (i)} 
a complete matter suppression of the self-induced transformations for
$t_{\rm pb}\lesssim 0.2$~s,
 {\em (ii)} 
partial matter suppression for $0.2\,\ \textrm{s}\lesssim t_{\rm pb}\lesssim 0.35$~s,
and
 {\em (iii)}
again complete suppression for $0.35\,\ \textrm{s} \lesssim t_{\rm pb}\lesssim 0.6$~s.
This feature suggests  a time-dependent pattern for the
$\nu$ conversions, i.e. complete-partial-complete suppression.

The behavior, analyzed for this specific example of the 10.8~M$_\odot$ SN
explosion model, is generic also for more massive iron-core SNe.
It is independent from the explosion scenarios and applies also for non-exploding
models. Indeed,   in any case the density of the material, enclosed inside the standing
bounce shock, can only increase due to mass accretion from the iron-core
envelope.
Only after the onset of an explosion, when mass accretion vanishes,
the matter density decreases.
However, for  the low-mass  O-Ne-Mg-core SNe, where the matter
 density profile is very steep, the suppression
 is  never complete.
As a consequence, the different features induced by the dense matter effects on the
oscillations may allow to distinguish iron-core SNe
from O-Ne-Mg-core SNe~\cite{Chakraborty:2011gd}.

Our results have been obtained considering a spherically symmetric neutrino emission. 
All the previous analysis in the field have relied on this assumption  to make
the flavor evolution equations numerically tractable.  It remains to be investigated if the 
removal of a perfect spherical symmetry can provide a different behavior in the flavor 
evolution~\cite{Sawyer:2008zs}.
Moreover, in multi-dimensional SN models density fluctuations are expected behind the standing 
bounce shock, due to the presence of convection and hydro instabilities. These can range  at 
most between 10$\%$ to a factor 2-3 (see, e.g.,~\cite{Tomas:2004gr,Scheck:2006rw}). Therefore,
even in this case, the matter suppression of the collective oscillations will still remain relevant. 
This claim is supported by a recent analysis of the matter suppression,  performed with two-dimensional
SN simulations~\cite{Dasgupta:2011jf}.

\begin{figure}[!t]
\begin{center}  
\includegraphics[width=1.\columnwidth]{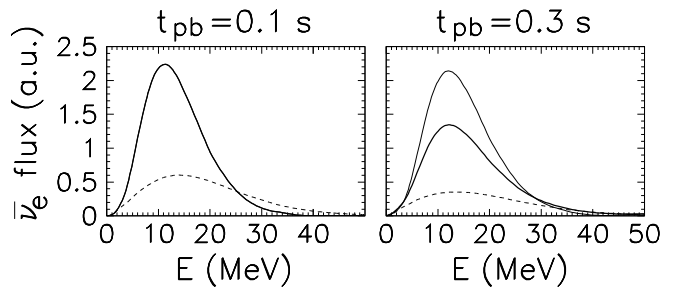}  
\end{center}  
\vspace{-0.5cm} 
\caption{Distribution functions for ${\bar\nu}_e$ at the neutrinosphere (continuous thin curve)
and after self-induced and matter effects at $r=2\times 10^3$~km (continuous thick 
curve), in the case of complete matter suppression ($P_{ee}=0$ at $t_{\rm pb}=0.1$~s, left panel) and
complete flavor mixture ($P_{ee}=1/2$ at $t_{\rm pb}=0.3$~s, right panel). 
For comparison, the oscillated ${\bar\nu}_e$ spectra obtained for
$n_e=0$ are also shown (dashed thin curve).
\label{fig4}} 
\end{figure}

\emph{Oscillated SN neutrino fluxes---}
Figure~\ref{fig4} shows the $\bar{\nu}_e$ distribution function  at the neutrinosphere (continuous thin curve)
as well as after self-induced and matter effects at $r=2\times 10^3$~km (continuous thick
curve).
We compare the case of complete matter suppression at $t_{\rm pb}=0.1$~s (left panel,
where thin and thick continuous curves coincide)
and complete flavor mixture at $t_{\rm pb}=0.3$~s (right panel).
We also show the oscillated flux for $n_e=0$, where a complete
$\bar{\nu}_e \to \bar{\nu}_x$ swap    occurs (dashed curve). 
The difference in the final  ${\bar\nu}_e$ flux with/without matter suppression
is striking. 
It is plausible that a high-statistics detection of a future  galactic
SN $\nu$ signal would monitor
the abrupt spectral changes between the phases of complete
and partial matter suppression, probing  this scenario.
These peculiar time variations in the $\nu$ signal during
the accretion would represent also a new tool to extract information on the
$\nu$ mass ordering, since the effects of dense matter would show up only
in the case of inverted mass hierarchy. 
%They may also be used to observe the exact time
%for the onset of the explosion.
 
 %%%%%%%%%%%%%%%%%%%%%%%%%%%%%%%%%%%%%%%%%%
%%%%%%%%%%%%%%%%%%%%%%%%%%%%%%%%%%%%%%%%%%%

 %%%%%%%%%%%%%%%%%%%%%%%%%%%%%%%%%%%%%%%%%%%
%%%%%%%%%%%%%%%%%%%%%%%%%%%%%%%%%%%%%%%%%%%
 
\emph{Earth matter effect.---}
A further consequence of the matter suppression is a significant change in the
interpretation of the Earth matter effect on the SN $\nu$ signal during the accretion
phase, occurring when $\nu$'s oscillate inside the Earth before being detected
(see, e.g., ~\cite{Lunardini:2001pb}). 
In the case of complete matter suppression of the self-induced oscillations
(at $t_{\rm pb}\lesssim 0.2$~s for iron-core SNe), 
the observable SN $\nu$ fluxes at Earth have been already calculated
in the  literature, antecedent to the inclusion of the collective effects. 
For definiteness, here we consider the Earth effects on the
$\bar{\nu}_e$ spectrum, observable through inverse beta decay reactions
$\bar{\nu}_e + p \to n + e^+$
at large volume  Cherenkov or scintillation detectors
(see, e.g.,~\cite{Dighe:2003jg}).

The ${\bar\nu}_e$ flux at Earth $F_{{\bar\nu}_e}^D$ in NH
 for any value of the mixing angle
 $\theta_{13}$  is  given by
%........................................................
$
F_{{\bar\nu}_e}^D = \cos^2 \theta_{12} F_{{\bar\nu}_e} + 
\sin^2 \theta_{12} F_{{\bar\nu}_x} 
$~\cite{Dighe:2003jg},
%.........................................................
where $\theta_{12}$ is the 1--2 mixing angle, with 
$\sin^2 \theta_{12}  \simeq 0.3$~\cite{Schwetz:2011qt}. 
In the IH case, for ``large'' $\theta_{13}$ (i.e. for $\sin^2\theta_{13}\gtrsim 10^{-3}$) 
$F_{{\bar\nu}_e}^D =F_{{\bar\nu}_x}$,
while  for  ``small''
$\theta_{13}$ (i.e. for $\sin^2\theta_{13}\lesssim 10^{-5}$)
the flux is the same as in the case of NH.
Earth effects can be taken into account by mapping
$\cos^2 \theta_{12} \to P({\bar\nu}_1 \to {\bar\nu}_e)$ and
$\sin^2 \theta_{12} \to 1-P({\bar\nu}_1 \to {\bar\nu}_e)$,
where $P({\bar\nu}_1 \to {\bar\nu}_e)$ is the probability that a state entering
the Earth as mass eigenstate ${\bar\nu}_1$ is detected as ${\bar\nu}_e$ at the
detector  (see, e.g.,~\cite{Lunardini:2001pb}).

In this scenario, the presence or absence of Earth matter effects at early times
($t_{\rm pb}\lesssim 0.2$~s)  will  allow to distinguish the
$\nu$ mass hierarchy at large value of the mixing angle $\theta_{13}$. 
This possibility,   prevented in the
previous scenario with dominant self-induced effects, 
is particularly attractive since there are already hints for a ``large''
$\theta_{13}$~\cite{Fogli:2008jx,:2011sj}, promising its possible detection with the
current and upcoming reactor and accelerator experiments~\cite{Mezzetto:2010zi}. 
Thus for large $\theta_{13}$, the next galactic SN $\nu$ signal would become
crucial to get a determination of the $\nu$ mass hierarchy from the sky. 
%The resolution of this ambiguity in the $\nu$ mass spectrum will
%shed
%light on how the lepton sector is organized and probe the underlying symmetries
%in the structure of $\nu$ masses and  mixing~\cite{Mohapatra:2006gs}.

\emph{Impact on SN heating.---}
Neutrino flavor oscillations between the neutrinospheres and the standing bounce
shock, have long been speculated to influence the $\nu$ heating and hence the
SN dynamics~\cite{shock,Duan:2006an}. 
In contrast, we find that the high matter density in the heating region causes 
complete  suppression  of the flavor conversion behind the shock-front.
Hence, collective $\nu$ flavor oscillations cannot help to increase the
$\nu$ heating significantly and an impact on the SN dynamics is not expected.
Our result is in agreement with the analysis recently performed in~\cite{Dasgupta:2011jf}. 
In order to solve the SN problem, which is related to the revival of the stalled
bounce shock, it is possible to decouple the $\nu$ flavor evolution from the
hydrodynamics aspects as well as from the $\nu$ transport.
%Nevertheless, $\nu$ flavor changes can have an impact on the 
%nucleosynthesis  during the SN cooling phase~\cite{Duan:2010af}. 

\emph{Conclusions.---}
In early, schematic investigations, the accretion phase seemed particularly promising to probe the
development of the collective $\nu$ transformations.
However, by analyzing state-of-the art simulations, we pointed out that the
presence of a large matter density
piled-up above the neutrinosphere can take its revenge over the
$\nu$--$\nu$ interactions, producing a significant suppression of the
self-induced flavor conversions. 
The presence of a large matter density during the accretion phase is 
a robust feature of SN
simulations~\cite{shock,Tomas:2004gr,Buras:2005rp, Liebendoerfer:2003es}.
However, its impact on  the self-induced oscillations was 
  estimated most often
negligible in previous studies (see, e.g.~\cite{Dasgupta:2009mg,Duan:2010bg}). 

Even if the matter suppression would prevent 
collective effects on the $\nu$ signal during the accretion phase, its presence will
result in various benefits. 
In particular, the detection of  the Earth matter effect on the SN $\nu$ burst during
the accretion  may  allow to extract the $\nu$ mass
hierarchy, if $\theta_{13}$ is not too small.
Moreover, the matter suppression of oscillations at high densities decouples the
problem of the $\nu$ flavor mixing in SNe from the $\nu$ transport and impact on
the  matter heating/cooling. 
   
Collective oscillations  may  remain possible during the
cooling phase, when the matter effects become sub-dominant due to the continuously
decreasing matter density.
However, the characterization of these effects in the presence of small flux differences
and of matter turbulences is far from being settled.
Further studies are crucial to understand possible effects of self-induced $\nu$
oscillations and imprinted observable signatures. 
%A possible future detection would provide an unique
%chance to probe the secret life of $\nu$'s in the deepest SN regions.

\vspace{-0.0cm}   

%%%%%%%%%%%%%%%%%%%%%%%%%%%%%%%%%%%%%%%%%%%%%%%%%%%%%%%%%%%%%%%%%%%%%%
\section*{Acknowledgements} %%%%%%%%%%%%%%%%%%%%%%%%%%%%%%%%%%%%%%%%%%%%%%%%
%%%%%%%%%%%%%%%%%%%%%%%%%%%%%%%%%%%%%%%%%%%%%%%%%%%%%%%%%%%%%%%%%%%%%%
We thank M.~Liebend\"{o}rfer,  E.~Lisi, C.~Ott, G.~Raffelt, S.~Sarikas, P.~D.~Serpico, G.~Sigl and I.~Tamborra for helpful comments on the manuscript. We also acknowledge  B.~Dasgupta and  T.~Janka for important discussions. 
The work of S.C., A.M., N.S.  was supported by the German Science Foundation (DFG)
within the Collaborative Research Center 676 ``Particles, Strings and the
Early Universe''. T.F. acknowledges support from HIC for FAIR project~no.~62800075.

\vspace{-0.5cm}   

%%%%%%%%%%%%%%%%%%%%%%%%%%%%%%%%%%%%%%%%%%%%%%%%%%%%%%%%%%%%%%%%%%%%%%
%\section*{References} %%%%%%%%%%%%%%%%%%%%%%%%%%%%%%%%%%%%%%%%%%%%%%%%
%%%%%%%%%%%%%%%%%%%%%%%%%%%%%%%%%%%%%%%%%%%%%%%%%%%%%%%%%%%%%%%%%%%%%%

%\vspace{-0.2cm}   

\end{document}